\documentclass[twocolumn,aps,prl,showpacs,amsmath,amssymb,nobibnotes]{revtex4-1}

\usepackage{graphicx}
\usepackage{amsmath,amssymb}
\usepackage{hyperref}

\begin{document}

\title{Reduction of trapped-ion anomalous heating by \textit{in situ} surface plasma cleaning}

\author{Robert McConnell}
\email{robert.mcconnell@ll.mit.edu}

\author{Colin Bruzewicz}
\email{colin.bruzewicz@ll.mit.edu}

\author{John Chiaverini}
\email{john.chiaverini@ll.mit.edu}

\author{Jeremy Sage}
\email{jsage@ll.mit.edu}
\affiliation{Lincoln Laboratory, Massachusetts Institute of Technology, Lexington, Massachusetts 02420, USA}

\date{\today}

\begin{abstract}
Anomalous motional heating is a major obstacle to scalable quantum information processing with trapped ions. While the source of this heating is not yet understood, several previous studies \cite{hite2012100,allcock2011reduction,chiaverini2014insensitivity} suggest that surface contaminants may be largely responsible. We demonstrate an improvement by a factor of four in the room-temperature heating rate of a niobium surface electrode trap by \textit{in situ} plasma cleaning of the trap surface. This surface treatment was performed with a simple homebuilt coil assembly and commercially-available matching network and is considerably gentler than other treatments, such as ion milling or laser cleaning, that have previously been shown to improve ion heating rates. We do not see an improvement in the heating rate when the trap is operated at cryogenic temperatures, pointing to a role of thermally-activated surface contaminants in motional heating whose activity may freeze out at low temperatures.
\end{abstract}

\maketitle

\section{Introduction}

Trapped ions form the basis of a promising technology for large-scale quantum information processing, combining very long coherence times with high-fidelity gate operations and scalable architectures. However, anomalous motional heating represents a major obstacle to be overcome before truly large-scale devices can be built \cite{turchette2000heating}.  This heating is called ``anomalous'' as it is orders of magnitude larger than known sources of heating, such as Johnson noise; its origins are currently not understood \cite{hiteMRS2013,brownnutt2014ion}. As all two-qubit gates demonstrated in trapped ions to date have utilized coupling between the motional and internal ion degrees of freedom, anomalous motional heating limits the achievable coherence and fidelity of two-qubit gates in ion traps. This heating has been found to increase strongly as the trapped ion is held closer to the electrode surface, making it a particularly important problem to be overcome if further miniaturization of ion traps is to continue. Available models suggest that this noise should be thermally activated, and significant reductions have been found by cooling ion traps to cryogenic temperatures \cite{PhysRevLett.97.103007,PhysRevLett.100.013001}, but even at low temperatures motional heating remains a significant limitation on achievable gate fidelity.

Several previous studies have pointed to the possible role of surface contaminants in producing anomalous heating. In Ref. \cite{chiaverini2014insensitivity}, the similar motional heating rates of two surface-electrode traps of the same geometry but different electrode material suggested that surface effects, rather than differences in the bulk, were responsible for the majority of the observed heating. Theoretical models have also been developed \cite{safavi2011microscopic, brownnutt2014ion} suggesting that surface adatoms or two-level fluctuators might produce electrical field noise that could give rise to the observed heating, although these models have so far failed to predict the detailed scaling behavior of ion-trap heating rates \cite{BruzewiczPRA2015}.

Furthermore, some previous experiments have shown an improvement in the heating rates of surface-electrode ion traps after surface treatments of the trap electrodes. Treatment with a high-energy pulsed laser source was shown to reduce a trap's heating rate by a factor of roughly three \cite{allcock2011reduction}. High-energy ion bombardment was observed to reduce trap heating rates by a factor of up to 100 \cite{hite2012100, daniilidis2014surface,mckay2014ion}, where it was verified that surface hydrocarbons were being removed by the process. The effectiveness of these treatments provides additional evidence that surface contaminants, particularly hydrocarbons, are likely a major contributor to ion motional heating. At the same time, treatment with high-energy laser pulses or ion beams can heat trap surfaces by hundreds of kelvin and produce additional undesirable effects: the trap in Ref. \cite{allcock2011reduction} showed visible damage in some locations due to laser heating, while keV-scale ion beams are known to sputter high-energy material from trap surfaces which can lead to unwanted metal redeposition. 

Radiofrequency (rf)-produced plasma is also known to be efficient at removing hydrocarbons from surfaces \cite{OKanePlasma1974} and is widely used to prepare surfaces for microfabrication processes and other applications. Plasma cleaning is a much gentler technique than pulsed laser cleaning or ion bombardment. Typical ion energies in an rf plasma are on the order of eV, such that sputtering of trap electrode material should be strongly suppressed \cite{YamamuraSputterYield1996}. Furthermore, rf plasma can be produced at relatively low rf power (in the range 5 - 20 W), so that an rf plasma source can be operated near a trap surface without excessive heating of the electrodes.

In this work, we report the use of \textit{in situ} rf plasma cleaning to reduce the room-temperature heating rate of a surface-electrode ion trap by a factor of four. We produce a mixed Ar-N$_{2}$-O$_{2}$ plasma with 15 W of rf power at 13.56 MHz coupled to a simple, home-wound coil which can be retracted after plasma cleaning to allow laser access and light collection for ion imaging without exposing the sample to air. Our method is comparatively gentle and heats the trap electrode surface by no more than about 25 K even after more than an hour of plasma cleaning. We also measure the ion trap heating rates at low temperature (4 K) and, interestingly, do not see an improvement from plasma cleaning. These results suggest that thermally-activated hydrocarbon contaminants play a significant role in anomalous motional heating of trapped ions, and that the activity of some (but possibly not all) of these contaminants freezes out at low temperatures and no longer causes heating.

\section{Experiment}

The ion-trapping apparatus used to perform these experiments has been extensively described elsewhere \cite{sage2012loading}. Briefly, we trap $^{88}$Sr$^{+}$ ions in a linear surface-electrode trap composed of Nb electrodes sputtered onto a sapphire substrate with typical metal thickness of 2 $\mu$m. A two-stage, vibrationally isolated cryocooler cools a low-temperature stage and an intermediate-temperature (50 K) shield which, along with a 50 l/s ion pump, provide ultra-high vacuum (UHV) conditions in the trap chamber without the need for an initial high-temperature bakeout. The trap chip itself is weakly coupled to the low-temperature stage, allowing the trap to be cooled to as low as 4 K; an on-chip heater allows us to heat the trap chip temperature to 295 K while the low-temperature stage remains below 10 K to retain effective cryopumping. A temperature sensor located adjacent to the trap chip indicates the trap chip temperature.

\begin{figure}
\includegraphics[width=0.5 \textwidth]{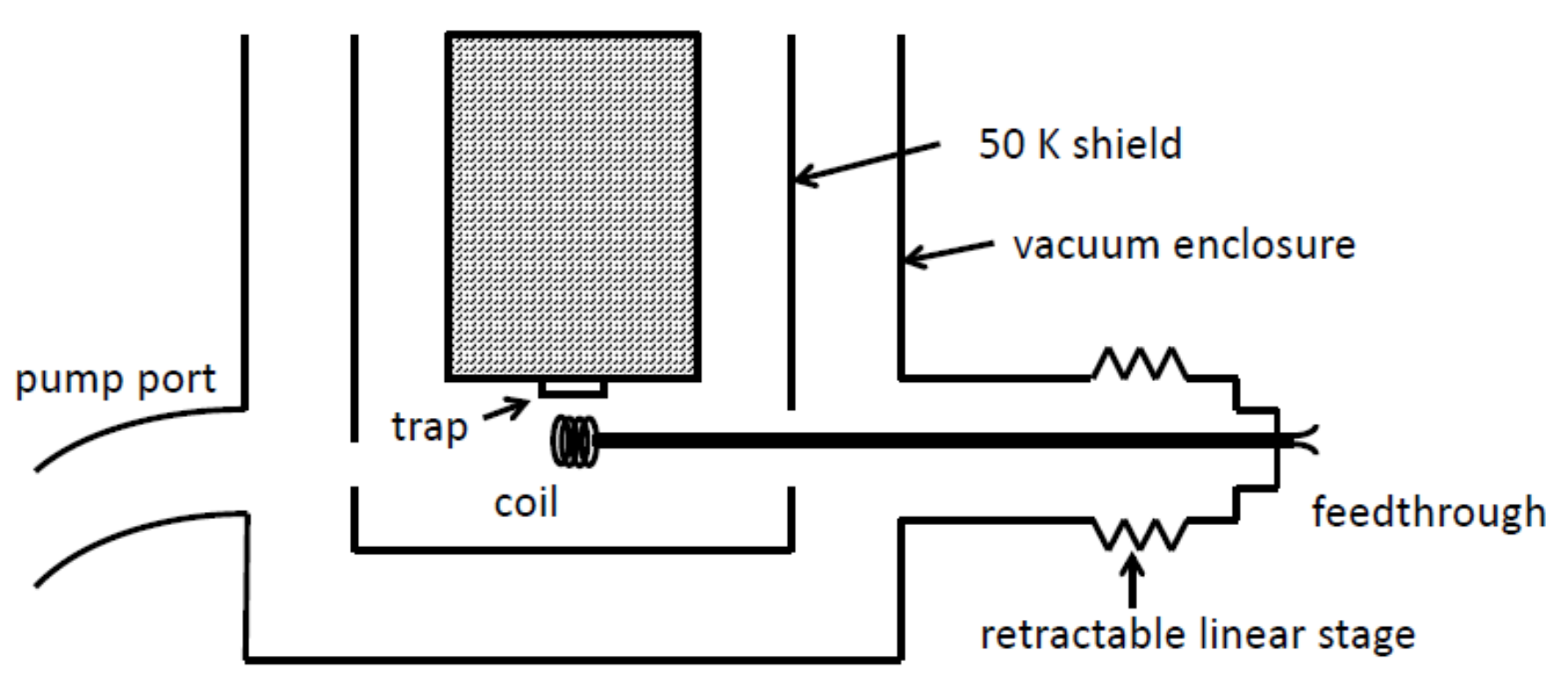}
\caption{Simplified schematic of the apparatus used for plasma cleaning studies. A surface-electrode ion trap on a temperature-controllable stage is enclosed within a 50 K radiation shield inside of a larger vacuum enclosure. The plasma is generated by rf power applied to a coil located near the ion trap. After plasma cleaning, a retractable linear stage allows the coil to be moved outside of the 50 K shield. Omitted from this simple schematic are many optical access ports as well as the source of neutral  $^{88}$Sr atoms. Figure not to scale.}
\label{Schematic}
\end{figure}

To load ions, we initially cool $^{88}$Sr atoms into a remotely-located magneto-optical trap (MOT), then use a resonant push beam to transfer atoms from the MOT to a region near the trap surface, where a pair of photoionization laser beams produce  $^{88}$Sr$^+$ ions. Those atoms which are ionized within the trapping volume can be confined, at a distance of 50 $\mu$m from the surface, with lifetimes on the order of minutes, due to the excellent cryogenic vacuum. We load a single ion which we then cool to the ground state of its axial motion (average vibrational occupation $\langle n \rangle < 0.1$) via Doppler cooling and resolved-sideband cooling. To measure heating rates, we apply a variable wait time after cooling the ion to its motional ground state and then measure the average occupation by the sideband-ratio technique \cite{PhysRevLett.75.4011}. 

Our rf plasma source consists of a 120 W, 13.56 MHz generator and impedance matching network (T\&C Power Conversion AG 0113 and AIT-600) coupled to a simple copper coil which is located near the trap chip. The coil consists of 6 turns of 22~AWG solid wire; the coil has diameter 1 cm, length 1.5 cm, and 25 cm-long leads. The coil is mounted on a standard 1.33" conflat feedthrough. Electrical shorts are prevented by passing the leads through rigid double-bore alumina tubing, which also provides mechanical stability. The entire coil assembly is mounted on a retractable linear shift stage (Kurt J. Lesker LSM38-100-H) which allows 10 cm of single-axis travel. The coil passes through a 1.5 cm diameter hole in the 50 K shield and is located about 1 cm vertically below our trap chip during plasma cleaning. A second hole of similar size in the opposite side of the 50 K shield allows the gas mixture to continually flow past the trap chip during plasma cleaning. The coil assembly then retracts out of the 50 K shield after plasma cleaning to allow laser and imaging access.

Our plasma cleaning procedure begins by pumping the system down to 50 mTorr with a roughing pump while at room temperature. We then introduce Ar gas at 300-400 mTorr into the system, while pumping to create a drift velocity of the background gas. We spark the plasma in a pure Ar environment with 15-20 W of rf power; we then reduce the rf power to 15 W and add gas from a $60 \% $ N$_{2}$ - $40\%$ O$_{2}$ mixture cylinder until the total system pressure is 700-800 mTorr, while maintaining plasma. This plasma is maintained for a variable length of time before the rf power is turned off, the system is pumped back down, and the cryocooler is turned on. Due to our cryogenic vacuum, we are able to reach UHV conditions (pressure $< 10^{-8}$ Torr) within about 3 h after plasma treatment without the need for a system bakeout.

In order to verify that our plasma cleaning technique actually removes surface hydrocarbons, we coated half of the surface of one of our Nb trap chips with a 1.5 $\mu$m-thick layer of a standard photoresist (AZ 1512) which is known to be removable by rf plasma. We then operated our plasma source for $\sim$ 60~min with parameters as described above. A Dektak contact profilometer was used to measure the height of the photoresist layer before and after plasma treatment. We found a reduction in surface height of 130 $\pm$ 10 nm, corresponding to a removal rate of about 2 nm / min. The removal appeared fairly uniform over the surface of the resist. In contrast, when we did not continually flow gas through the chamber during plasma cleaning, we also saw material removal, but the removal was extremely uneven across the surface, leading to the possibility that some regions of the chip would not be cleaned effectively.

To characterize the effects of plasma cleaning on ion motional heating, we used our plasma source to clean two identical Nb surface-electrode traps, which we designate Trap A and Trap B. We ran the plasma source for variable lengths of time, but with parameters otherwise as described above. We compared motional heating rates before and after plasma cleaning in both traps. We measured at two trap electrode temperatures (295 K and 4 K), as well as two axial trap frequencies (660 kHz and 1.3 MHz). We conducted additional tests on Trap A to further characterize the heating rate. 

\section{Results}

Figure 2 shows the ion motional heating rate in quanta/s for Trap A, before (red squares) and after (black circles) 20 min of plasma cleaning. We find a reduction by approximately a factor of two in the trap heating rate at room temperature after this treatment, at both axial trap frequencies investigated. However, the heating rate when the trap chip is held at 4 K is not significantly improved by the plasma cleaning. The frequency dependence of the heating rate is similar to what we have seen in previous measurements of traps with the same geometry \cite{BruzewiczPRA2015}, and is not changed by the plasma cleaning.

\begin{figure}
\includegraphics[width=0.5 \textwidth]{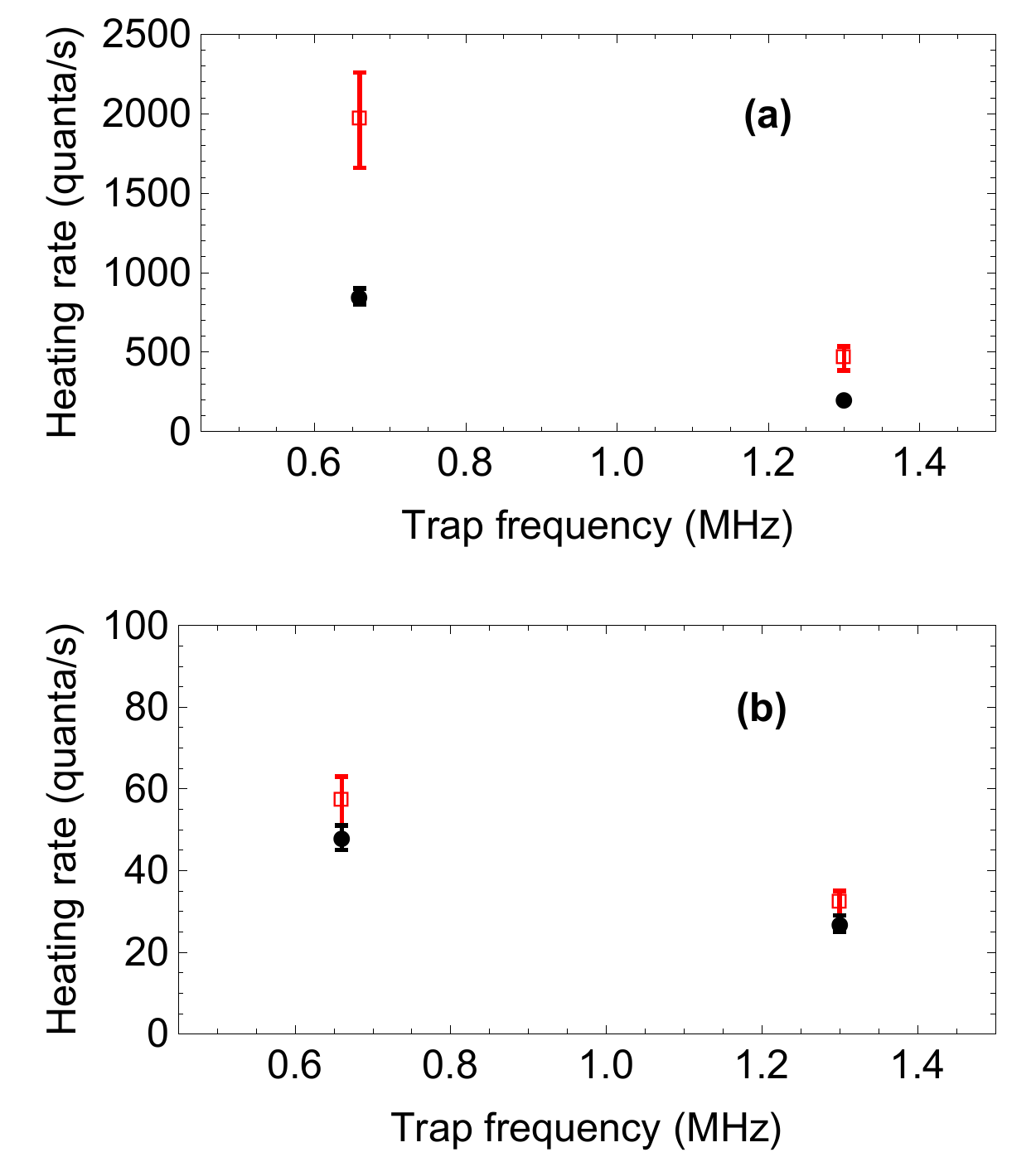}
\caption{Ion heating rate in Trap A before (open red squares) and after (filled black circles) 20-minute plasma cleaning, as a function of trap frequency, at (a) $T = 295$ K and (b) $T  = 4$ K trap chip temperature.}
\label{PlasmaHeatingRates}
\end{figure}

To ensure that the observed reduction in heating rate is due to the plasma treatment, we vented Trap A to air for 72 h, then repeated our sequence of measurements. After this air exposure we found that the trap's room-temperature heating rates increased from their post-plasma values, but did not quite return to their initial values. We then applied a second, 20-minute plasma cleaning, after which the trap heating rates decreased even further, to only 25--30\% of their initial values. This motivated us to try a very long, 75-minute plasma cleaning on Trap A. However, we did not see further improvement as a result of this treatment, suggesting that we had reached the limits of heating-rate reduction achievable with the current procedure. The time schedule of plasma cleanings and air exposures, with their associated room-temperature heating rates, is shown in Figure 3. We note that at no point did we see an improvement of the low-temperature trap heating rate due to plasma cleaning. During our 75-minute plasma cleaning step (the longest used in these experiments), the temperature as measured by the sensor near the trap chip increased by only 24 K. 

For reference, Trap A's heating rate at 1.3 MHz trap frequency and 295 K before plasma cleaning corresponds to electric-field noise spectral density of $S_E (f) = 9.0 \times 10^{-12}$ (V/m)$^2$/Hz, which ultimately decreased to a final value of  $S_E (f) = 2.4 \times 10^{-12}$(V/m)$^2$/Hz after all plasma cleaning steps. Even before plasma-cleaning, this heating rate compares favorably to rates seen in other ion traps with similar geometry \cite{chiaverini2014insensitivity}.

\begin{figure}
\includegraphics[width=0.5 \textwidth]{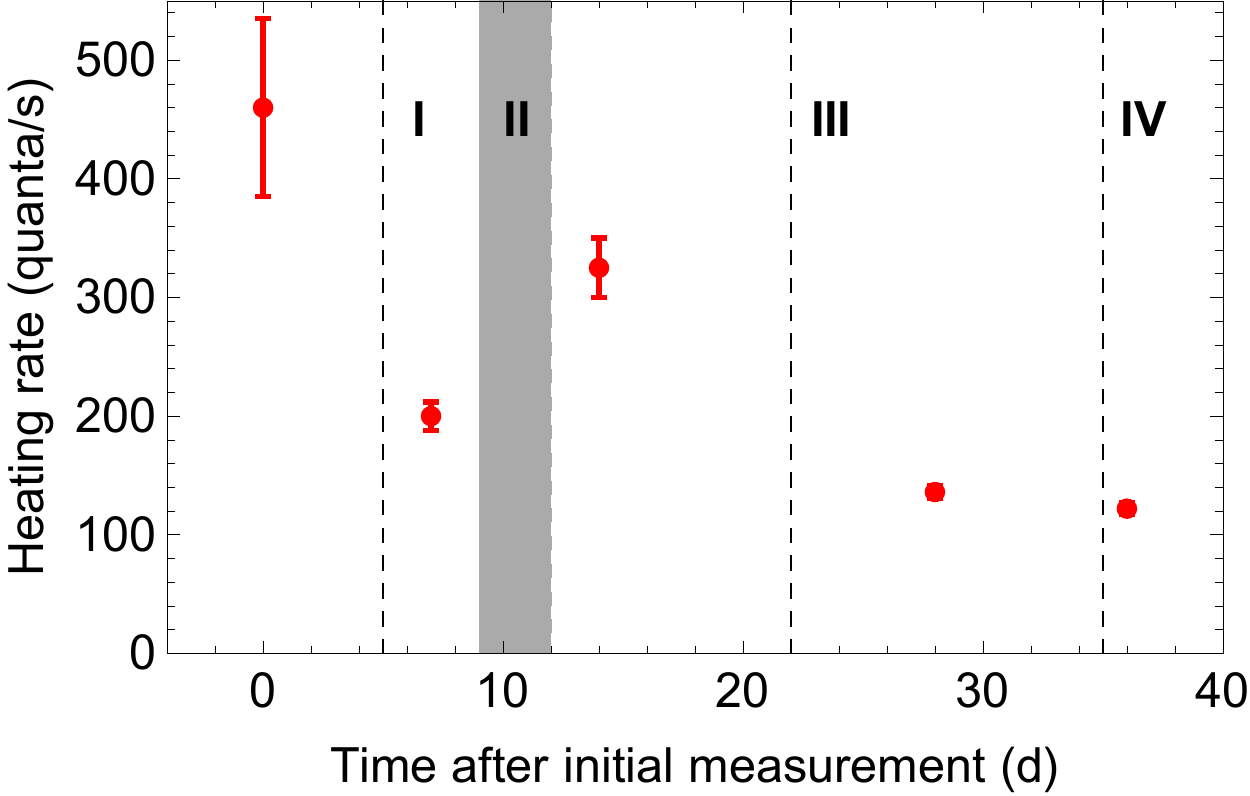}
\caption{Time course of the Trap A heating rate (for chip temperature of 295 K and trap frequency of 1.3 MHz). Vertical dashed lines show plasma cleanings (I and III for 20 minutes, IV for 75 minutes), while the gray area (II) indicates the 72-h exposure to air to allow surface contaminants to resorb to the trap.}
\label{PlasmaTimePlot}
\end{figure}

Finally, to further assess the repeatability of our treatment, we applied plasma cleaning to a second trap, Trap B, identical in design to Trap A. After conducting an initial series of heating rate measurements on Trap B, we applied a 75-minute plasma cleaning sequence. We chose to use this long cleaning time as we had observed, in Trap A, that a cleaning time of only 20 min was not sufficient to reach the lowest possible heating rates. After applying the plasma cleaning process and pumping our chamber down to UHV conditions, we then waited for 120 h before initiating measurements, allowing us to verify that reduction in heating rates can last at least several days. Figure 4 shows the room-temperature and cryogenic heating rates of Trap B before and after plasma cleaning for 75 minutes. This single plasma cleaning step resulted in a heating rate improvement at room temperature of a factor of $3.1 \pm 0.6$ at 660 kHz trap frequency and a factor of $3.8 \pm 0.3$ at 1.3 MHz trap frequency.

\begin{figure}[t!]
\includegraphics[width=0.5 \textwidth]{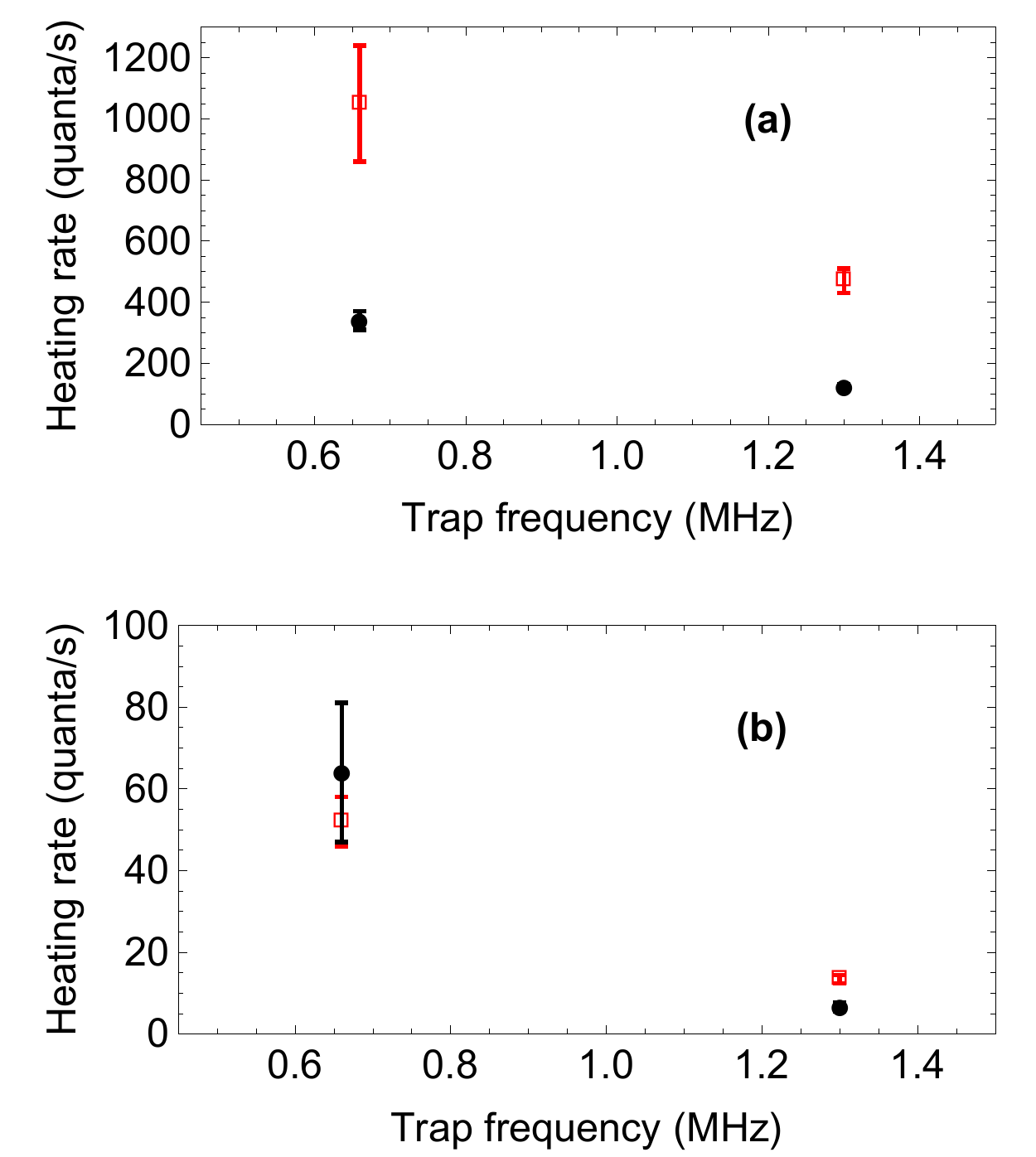}
\caption{Ion heating rate in Trap B before (open red squares) and after (filled black circles) 75-minute plasma cleaning, as a function of trap frequency at (a) $T = 295$ K and (b) $T  =4$ K trap chip temperature.}
\label{PlasmaHeatingRatesNewTrap}
\end{figure}

For Trap B, cryogenic measurements did not indicate any improvement in heating rate at 660 kHz, but did indicate a small but significant improvement in heating rate at 1.3 MHz. Any improvement at low temperature is clearly much less dramatic than the improvement at 295 K, consistent with the results observed in Trap A. The Trap B data at temperature of 4 K and trap frequency of 1.3 MHz are close to the lowest heating rates we have observed in this apparatus, so we cannot rule out the possibility that this particular set of measurements is limited by technical noise. The observation of such low heating rates under these conditions, however, offers strong evidence that the other sets of heating rate measurements are not limited by technical noise.

\section{Conclusions}

In conclusion, we have demonstrated a new technique to reduce the anomalous motional heating of trapped ions: \textit{in situ} plasma cleaning. This approach is simple and robust and causes minimal perturbation to the trap electrode material, unlike other surface treatments which have previously demonstrated reduction of ion motional heating. We have demonstrated a reduction by a factor of three to four in the room-temperature heating rate via a 75-minute, low-power rf plasma cleaning. Interestingly, we did not observe a similar reduction in the heating rate when the trap electrodes were at cryogenic temperatures, possibly indicating that the plasma's role is to remove thermally-activated surface contaminants which are frozen out at low temperature.

It is possible that the lowest achievable heating rates in ion traps will combine multiple surface treatments. For example, there is some evidence that high-energy ion-bombardment techniques cause some structural reorganization of the trap electrode material at the surface \cite{HiteChat}, which may be in part responsible for the reduction in heating rates associated with this technique. It is possible that a one-time, \textit{ex situ} ion bombardment may initially lower the heating rate, while periodic \textit{in situ} plasma cleaning may be able to remove contaminants that slowly adsorb onto the trap surface, especially if a system bakeout is not required after cleaning.

\section{Acknowledgements}

We thank Peter Murphy, Chris Thoummaraj, and Karen Magoon for assistance with ion trap chip packaging. We thank Jeanne Porter for resist coating of samples and profilometer assistance. We thank Steven Vitale for helpful discussions and advice. This work was sponsored by the Assistant Secretary of Defense for Research and Engineering under Air Force Contract \#FA8721-05-C-0002. Opinions, interpretations, conclusions, and recommendations are those of the authors and are not necessarily endorsed by the United States Government.


\begin{thebibliography}{17}%
\makeatletter
\providecommand \@ifxundefined [1]{%
 \@ifx{#1\undefined}
}%
\providecommand \@ifnum [1]{%
 \ifnum #1\expandafter \@firstoftwo
 \else \expandafter \@secondoftwo
 \fi
}%
\providecommand \@ifx [1]{%
 \ifx #1\expandafter \@firstoftwo
 \else \expandafter \@secondoftwo
 \fi
}%
\providecommand \natexlab [1]{#1}%
\providecommand \enquote  [1]{``#1''}%
\providecommand \bibnamefont  [1]{#1}%
\providecommand \bibfnamefont [1]{#1}%
\providecommand \citenamefont [1]{#1}%
\providecommand \href@noop [0]{\@secondoftwo}%
\providecommand \href [0]{\begingroup \@sanitize@url \@href}%
\providecommand \@href[1]{\@@startlink{#1}\@@href}%
\providecommand \@@href[1]{\endgroup#1\@@endlink}%
\providecommand \@sanitize@url [0]{\catcode `\\12\catcode `\$12\catcode
  `\&12\catcode `\#12\catcode `\^12\catcode `\_12\catcode `\%12\relax}%
\providecommand \@@startlink[1]{}%
\providecommand \@@endlink[0]{}%
\providecommand \url  [0]{\begingroup\@sanitize@url \@url }%
\providecommand \@url [1]{\endgroup\@href {#1}{\urlprefix }}%
\providecommand \urlprefix  [0]{URL }%
\providecommand \Eprint [0]{\href }%
\providecommand \doibase [0]{http://dx.doi.org/}%
\providecommand \selectlanguage [0]{\@gobble}%
\providecommand \bibinfo  [0]{\@secondoftwo}%
\providecommand \bibfield  [0]{\@secondoftwo}%
\providecommand \translation [1]{[#1]}%
\providecommand \BibitemOpen [0]{}%
\providecommand \bibitemStop [0]{}%
\providecommand \bibitemNoStop [0]{.\EOS\space}%
\providecommand \EOS [0]{\spacefactor3000\relax}%
\providecommand \BibitemShut  [1]{\csname bibitem#1\endcsname}%
\let\auto@bib@innerbib\@empty
\bibitem [{\citenamefont {Hite}\ \emph {et~al.}(2012)\citenamefont {Hite},
  \citenamefont {Colombe}, \citenamefont {Wilson}, \citenamefont {Brown},
  \citenamefont {Warring}, \citenamefont {J{\"o}rdens}, \citenamefont {Jost},
  \citenamefont {McKay}, \citenamefont {Pappas}, \citenamefont {Leibfried},\
  and\ \citenamefont {Wineland}}]{hite2012100}%
  \BibitemOpen
  \bibfield  {author} {\bibinfo {author} {\bibfnamefont {D.~A.}\ \bibnamefont
  {Hite}}, \bibinfo {author} {\bibfnamefont {Y.}~\bibnamefont {Colombe}},
  \bibinfo {author} {\bibfnamefont {A.~C.}\ \bibnamefont {Wilson}}, \bibinfo
  {author} {\bibfnamefont {K.~R.}\ \bibnamefont {Brown}}, \bibinfo {author}
  {\bibfnamefont {U.}~\bibnamefont {Warring}}, \bibinfo {author} {\bibfnamefont
  {R.}~\bibnamefont {J{\"o}rdens}}, \bibinfo {author} {\bibfnamefont {J.~D.}\
  \bibnamefont {Jost}}, \bibinfo {author} {\bibfnamefont {K.~S.}\ \bibnamefont
  {McKay}}, \bibinfo {author} {\bibfnamefont {D.~P.}\ \bibnamefont {Pappas}},
  \bibinfo {author} {\bibfnamefont {D.}~\bibnamefont {Leibfried}}, \ and\
  \bibinfo {author} {\bibfnamefont {D.~J.}\ \bibnamefont {Wineland}},\
  }\href@noop {} {\bibfield  {journal} {\bibinfo  {journal} {Phys. Rev. Lett.}\
  }\textbf {\bibinfo {volume} {109}},\ \bibinfo {pages} {103001} (\bibinfo
  {year} {2012})}\BibitemShut {NoStop}%
\bibitem [{\citenamefont {Allcock}\ \emph {et~al.}(2011)\citenamefont
  {Allcock}, \citenamefont {Guidoni}, \citenamefont {Harty}, \citenamefont
  {Ballance}, \citenamefont {Blain}, \citenamefont {Steane},\ and\
  \citenamefont {Lucas}}]{allcock2011reduction}%
  \BibitemOpen
  \bibfield  {author} {\bibinfo {author} {\bibfnamefont {D.~T.~C.}\
  \bibnamefont {Allcock}}, \bibinfo {author} {\bibfnamefont {L.}~\bibnamefont
  {Guidoni}}, \bibinfo {author} {\bibfnamefont {T.~P.}\ \bibnamefont {Harty}},
  \bibinfo {author} {\bibfnamefont {C.~J.}\ \bibnamefont {Ballance}}, \bibinfo
  {author} {\bibfnamefont {M.~G.}\ \bibnamefont {Blain}}, \bibinfo {author}
  {\bibfnamefont {A.~M.}\ \bibnamefont {Steane}}, \ and\ \bibinfo {author}
  {\bibfnamefont {D.~M.}\ \bibnamefont {Lucas}},\ }\href@noop {} {\bibfield
  {journal} {\bibinfo  {journal} {New J. Phys.}\ }\textbf {\bibinfo {volume}
  {13}},\ \bibinfo {pages} {123023} (\bibinfo {year} {2011})}\BibitemShut
  {NoStop}%
\bibitem [{\citenamefont {Chiaverini}\ and\ \citenamefont
  {Sage}(2014)}]{chiaverini2014insensitivity}%
  \BibitemOpen
  \bibfield  {author} {\bibinfo {author} {\bibfnamefont {J.}~\bibnamefont
  {Chiaverini}}\ and\ \bibinfo {author} {\bibfnamefont {J.~M.}\ \bibnamefont
  {Sage}},\ }\href@noop {} {\bibfield  {journal} {\bibinfo  {journal} {Phys.
  Rev. A}\ }\textbf {\bibinfo {volume} {89}},\ \bibinfo {pages} {012318}
  (\bibinfo {year} {2014})}\BibitemShut {NoStop}%
\bibitem [{\citenamefont {Turchette}\ \emph {et~al.}(2000)\citenamefont
  {Turchette}, \citenamefont {Kielpinski}, \citenamefont {King}, \citenamefont
  {Leibfried}, \citenamefont {Meekhof}, \citenamefont {Myatt}, \citenamefont
  {Rowe}, \citenamefont {Sackett}, \citenamefont {Wood}, \citenamefont {Itano},
  \citenamefont {Monroe},\ and\ \citenamefont
  {Wineland}}]{turchette2000heating}%
  \BibitemOpen
  \bibfield  {author} {\bibinfo {author} {\bibfnamefont {Q.~A.}\ \bibnamefont
  {Turchette}}, \bibinfo {author} {\bibfnamefont {D.}~\bibnamefont
  {Kielpinski}}, \bibinfo {author} {\bibfnamefont {B.~E.}\ \bibnamefont
  {King}}, \bibinfo {author} {\bibfnamefont {D.}~\bibnamefont {Leibfried}},
  \bibinfo {author} {\bibfnamefont {D.~M.}\ \bibnamefont {Meekhof}}, \bibinfo
  {author} {\bibfnamefont {C.~J.}\ \bibnamefont {Myatt}}, \bibinfo {author}
  {\bibfnamefont {M.~A.}\ \bibnamefont {Rowe}}, \bibinfo {author}
  {\bibfnamefont {C.~A.}\ \bibnamefont {Sackett}}, \bibinfo {author}
  {\bibfnamefont {C.~S.}\ \bibnamefont {Wood}}, \bibinfo {author}
  {\bibfnamefont {W.}~\bibnamefont {Itano}}, \bibinfo {author} {\bibfnamefont
  {C.}~\bibnamefont {Monroe}}, \ and\ \bibinfo {author} {\bibfnamefont {D.~J.}\
  \bibnamefont {Wineland}},\ }\href@noop {} {\bibfield  {journal} {\bibinfo
  {journal} {Phys. Rev. A}\ }\textbf {\bibinfo {volume} {61}},\ \bibinfo
  {pages} {063418} (\bibinfo {year} {2000})}\BibitemShut {NoStop}%
\bibitem [{\citenamefont {Hite}\ \emph {et~al.}(2013)\citenamefont {Hite},
  \citenamefont {Colombe}, \citenamefont {Wilson}, \citenamefont {Allcock},
  \citenamefont {Leibried}, \citenamefont {Wineland},\ and\ \citenamefont
  {Pappas}}]{hiteMRS2013}%
  \BibitemOpen
  \bibfield  {author} {\bibinfo {author} {\bibfnamefont {D.~A.}\ \bibnamefont
  {Hite}}, \bibinfo {author} {\bibfnamefont {Y.}~\bibnamefont {Colombe}},
  \bibinfo {author} {\bibfnamefont {A.~C.}\ \bibnamefont {Wilson}}, \bibinfo
  {author} {\bibfnamefont {D.~T.~C.}\ \bibnamefont {Allcock}}, \bibinfo
  {author} {\bibfnamefont {D.}~\bibnamefont {Leibried}}, \bibinfo {author}
  {\bibfnamefont {D.~J.}\ \bibnamefont {Wineland}}, \ and\ \bibinfo {author}
  {\bibfnamefont {D.~P.}\ \bibnamefont {Pappas}},\ }\href@noop {} {\bibfield
  {journal} {\bibinfo  {journal} {MRS Bulletin}\ }\textbf {\bibinfo {volume}
  {38}},\ \bibinfo {pages} {826} (\bibinfo {year} {2013})}\BibitemShut
  {NoStop}%
\bibitem [{\citenamefont {Brownnutt}\ \emph {et~al.}(2014)\citenamefont
  {Brownnutt}, \citenamefont {Kumph}, \citenamefont {Rabl},\ and\ \citenamefont
  {Blatt}}]{brownnutt2014ion}%
  \BibitemOpen
  \bibfield  {author} {\bibinfo {author} {\bibfnamefont {M.}~\bibnamefont
  {Brownnutt}}, \bibinfo {author} {\bibfnamefont {M.}~\bibnamefont {Kumph}},
  \bibinfo {author} {\bibfnamefont {P.}~\bibnamefont {Rabl}}, \ and\ \bibinfo
  {author} {\bibfnamefont {R.}~\bibnamefont {Blatt}},\ }\href@noop {}
  {\bibfield  {journal} {\bibinfo  {journal} {arXiv preprint arXiv:1409.6572}\
  } (\bibinfo {year} {2014})}\BibitemShut {NoStop}%
\bibitem [{\citenamefont {Deslauriers}\ \emph {et~al.}(2006)\citenamefont
  {Deslauriers}, \citenamefont {Olmschenk}, \citenamefont {Stick},
  \citenamefont {Hensinger}, \citenamefont {Sterk},\ and\ \citenamefont
  {Monroe}}]{PhysRevLett.97.103007}%
  \BibitemOpen
  \bibfield  {author} {\bibinfo {author} {\bibfnamefont {L.}~\bibnamefont
  {Deslauriers}}, \bibinfo {author} {\bibfnamefont {S.}~\bibnamefont
  {Olmschenk}}, \bibinfo {author} {\bibfnamefont {D.}~\bibnamefont {Stick}},
  \bibinfo {author} {\bibfnamefont {W.~K.}\ \bibnamefont {Hensinger}}, \bibinfo
  {author} {\bibfnamefont {J.}~\bibnamefont {Sterk}}, \ and\ \bibinfo {author}
  {\bibfnamefont {C.}~\bibnamefont {Monroe}},\ }\href {\doibase
  10.1103/PhysRevLett.97.103007} {\bibfield  {journal} {\bibinfo  {journal}
  {Phys. Rev. Lett.}\ }\textbf {\bibinfo {volume} {97}},\ \bibinfo {pages}
  {103007} (\bibinfo {year} {2006})}\BibitemShut {NoStop}%
\bibitem [{\citenamefont {Labaziewicz}\ \emph {et~al.}(2008)\citenamefont
  {Labaziewicz}, \citenamefont {Ge}, \citenamefont {Antohi}, \citenamefont
  {Leibrandt}, \citenamefont {Brown},\ and\ \citenamefont
  {Chuang}}]{PhysRevLett.100.013001}%
  \BibitemOpen
  \bibfield  {author} {\bibinfo {author} {\bibfnamefont {J.}~\bibnamefont
  {Labaziewicz}}, \bibinfo {author} {\bibfnamefont {Y.}~\bibnamefont {Ge}},
  \bibinfo {author} {\bibfnamefont {P.}~\bibnamefont {Antohi}}, \bibinfo
  {author} {\bibfnamefont {D.}~\bibnamefont {Leibrandt}}, \bibinfo {author}
  {\bibfnamefont {K.~R.}\ \bibnamefont {Brown}}, \ and\ \bibinfo {author}
  {\bibfnamefont {I.~L.}\ \bibnamefont {Chuang}},\ }\href {\doibase
  10.1103/PhysRevLett.100.013001} {\bibfield  {journal} {\bibinfo  {journal}
  {Phys. Rev. Lett.}\ }\textbf {\bibinfo {volume} {100}},\ \bibinfo {pages}
  {013001} (\bibinfo {year} {2008})}\BibitemShut {NoStop}%
\bibitem [{\citenamefont {Safavi-Naini}\ \emph {et~al.}(2011)\citenamefont
  {Safavi-Naini}, \citenamefont {Rabl}, \citenamefont {Weck},\ and\
  \citenamefont {Sadeghpour}}]{safavi2011microscopic}%
  \BibitemOpen
  \bibfield  {author} {\bibinfo {author} {\bibfnamefont {A.}~\bibnamefont
  {Safavi-Naini}}, \bibinfo {author} {\bibfnamefont {P.}~\bibnamefont {Rabl}},
  \bibinfo {author} {\bibfnamefont {P.~F.}\ \bibnamefont {Weck}}, \ and\
  \bibinfo {author} {\bibfnamefont {H.~R.}\ \bibnamefont {Sadeghpour}},\
  }\href@noop {} {\bibfield  {journal} {\bibinfo  {journal} {Phys. Rev. A}\
  }\textbf {\bibinfo {volume} {84}},\ \bibinfo {pages} {023412} (\bibinfo
  {year} {2011})}\BibitemShut {NoStop}%
\bibitem [{\citenamefont {Bruzewicz}\ \emph {et~al.}(2015)\citenamefont
  {Bruzewicz}, \citenamefont {Sage},\ and\ \citenamefont
  {Chiaverini}}]{BruzewiczPRA2015}%
  \BibitemOpen
  \bibfield  {author} {\bibinfo {author} {\bibfnamefont {C.~D.}\ \bibnamefont
  {Bruzewicz}}, \bibinfo {author} {\bibfnamefont {J.~M.}\ \bibnamefont {Sage}},
  \ and\ \bibinfo {author} {\bibfnamefont {J.}~\bibnamefont {Chiaverini}},\
  }\href {\doibase 10.1103/PhysRevA.91.041402} {\bibfield  {journal} {\bibinfo
  {journal} {Phys. Rev. A}\ }\textbf {\bibinfo {volume} {91}},\ \bibinfo
  {pages} {041402} (\bibinfo {year} {2015})}\BibitemShut {NoStop}%
\bibitem [{\citenamefont {Daniilidis}\ \emph {et~al.}(2014)\citenamefont
  {Daniilidis}, \citenamefont {Gerber}, \citenamefont {Bolloten}, \citenamefont
  {Ramm}, \citenamefont {Ransford}, \citenamefont {Ulin-Avila}, \citenamefont
  {Talukdar},\ and\ \citenamefont {H{\"a}ffner}}]{daniilidis2014surface}%
  \BibitemOpen
  \bibfield  {author} {\bibinfo {author} {\bibfnamefont {N.}~\bibnamefont
  {Daniilidis}}, \bibinfo {author} {\bibfnamefont {S.}~\bibnamefont {Gerber}},
  \bibinfo {author} {\bibfnamefont {G.}~\bibnamefont {Bolloten}}, \bibinfo
  {author} {\bibfnamefont {M.}~\bibnamefont {Ramm}}, \bibinfo {author}
  {\bibfnamefont {A.}~\bibnamefont {Ransford}}, \bibinfo {author}
  {\bibfnamefont {E.}~\bibnamefont {Ulin-Avila}}, \bibinfo {author}
  {\bibfnamefont {I.}~\bibnamefont {Talukdar}}, \ and\ \bibinfo {author}
  {\bibfnamefont {H.}~\bibnamefont {H{\"a}ffner}},\ }\href@noop {} {\bibfield
  {journal} {\bibinfo  {journal} {Phys. Rev. B}\ }\textbf {\bibinfo {volume}
  {89}},\ \bibinfo {pages} {245435} (\bibinfo {year} {2014})}\BibitemShut
  {NoStop}%
\bibitem [{\citenamefont {McKay}\ \emph {et~al.}(2014)\citenamefont {McKay},
  \citenamefont {Hite}, \citenamefont {Colombe}, \citenamefont {J{\"o}rdens},
  \citenamefont {Wilson}, \citenamefont {Slichter}, \citenamefont {Allcock},
  \citenamefont {Leibfried}, \citenamefont {Wineland},\ and\ \citenamefont
  {Pappas}}]{mckay2014ion}%
  \BibitemOpen
  \bibfield  {author} {\bibinfo {author} {\bibfnamefont {K.~S.}\ \bibnamefont
  {McKay}}, \bibinfo {author} {\bibfnamefont {D.~A.}\ \bibnamefont {Hite}},
  \bibinfo {author} {\bibfnamefont {Y.}~\bibnamefont {Colombe}}, \bibinfo
  {author} {\bibfnamefont {R.}~\bibnamefont {J{\"o}rdens}}, \bibinfo {author}
  {\bibfnamefont {A.~C.}\ \bibnamefont {Wilson}}, \bibinfo {author}
  {\bibfnamefont {D.~H.}\ \bibnamefont {Slichter}}, \bibinfo {author}
  {\bibfnamefont {D.~T.~C.}\ \bibnamefont {Allcock}}, \bibinfo {author}
  {\bibfnamefont {D.}~\bibnamefont {Leibfried}}, \bibinfo {author}
  {\bibfnamefont {D.~J.}\ \bibnamefont {Wineland}}, \ and\ \bibinfo {author}
  {\bibfnamefont {D.~P.}\ \bibnamefont {Pappas}},\ }\href@noop {} {\bibfield
  {journal} {\bibinfo  {journal} {arXiv preprint arXiv:1406.1778}\ } (\bibinfo
  {year} {2014})}\BibitemShut {NoStop}%
\bibitem [{\citenamefont {O'Kane}\ and\ \citenamefont
  {Mittal}(1974)}]{OKanePlasma1974}%
  \BibitemOpen
  \bibfield  {author} {\bibinfo {author} {\bibfnamefont {D.~F.}\ \bibnamefont
  {O'Kane}}\ and\ \bibinfo {author} {\bibfnamefont {K.~L.}\ \bibnamefont
  {Mittal}},\ }\href@noop {} {\bibfield  {journal} {\bibinfo  {journal} {Vacuum
  Science and Technology}\ }\textbf {\bibinfo {volume} {11}},\ \bibinfo {pages}
  {567} (\bibinfo {year} {1974})}\BibitemShut {NoStop}%
\bibitem [{\citenamefont {Yamamura}\ and\ \citenamefont
  {Tawara}(1996)}]{YamamuraSputterYield1996}%
  \BibitemOpen
  \bibfield  {author} {\bibinfo {author} {\bibfnamefont {Y.}~\bibnamefont
  {Yamamura}}\ and\ \bibinfo {author} {\bibfnamefont {H.}~\bibnamefont
  {Tawara}},\ }\href {\doibase http://dx.doi.org/10.1006/adnd.1996.0005}
  {\bibfield  {journal} {\bibinfo  {journal} {Atomic Data and Nuclear Data
  Tables}\ }\textbf {\bibinfo {volume} {62}},\ \bibinfo {pages} {149 }
  (\bibinfo {year} {1996})}\BibitemShut {NoStop}%
\bibitem [{\citenamefont {Sage}\ \emph {et~al.}(2012)\citenamefont {Sage},
  \citenamefont {Kerman},\ and\ \citenamefont {Chiaverini}}]{sage2012loading}%
  \BibitemOpen
  \bibfield  {author} {\bibinfo {author} {\bibfnamefont {J.~M.}\ \bibnamefont
  {Sage}}, \bibinfo {author} {\bibfnamefont {A.~J.}\ \bibnamefont {Kerman}}, \
  and\ \bibinfo {author} {\bibfnamefont {J.}~\bibnamefont {Chiaverini}},\
  }\href@noop {} {\bibfield  {journal} {\bibinfo  {journal} {Phys. Rev. A}\
  }\textbf {\bibinfo {volume} {86}},\ \bibinfo {pages} {013417} (\bibinfo
  {year} {2012})}\BibitemShut {NoStop}%
\bibitem [{\citenamefont {Monroe}\ \emph {et~al.}(1995)\citenamefont {Monroe},
  \citenamefont {Meekhof}, \citenamefont {King}, \citenamefont {Jefferts},
  \citenamefont {Itano}, \citenamefont {Wineland},\ and\ \citenamefont
  {Gould}}]{PhysRevLett.75.4011}%
  \BibitemOpen
  \bibfield  {author} {\bibinfo {author} {\bibfnamefont {C.}~\bibnamefont
  {Monroe}}, \bibinfo {author} {\bibfnamefont {D.~M.}\ \bibnamefont {Meekhof}},
  \bibinfo {author} {\bibfnamefont {B.~E.}\ \bibnamefont {King}}, \bibinfo
  {author} {\bibfnamefont {S.~R.}\ \bibnamefont {Jefferts}}, \bibinfo {author}
  {\bibfnamefont {W.~M.}\ \bibnamefont {Itano}}, \bibinfo {author}
  {\bibfnamefont {D.~J.}\ \bibnamefont {Wineland}}, \ and\ \bibinfo {author}
  {\bibfnamefont {P.}~\bibnamefont {Gould}},\ }\href {\doibase
  10.1103/PhysRevLett.75.4011} {\bibfield  {journal} {\bibinfo  {journal}
  {Phys. Rev. Lett.}\ }\textbf {\bibinfo {volume} {75}},\ \bibinfo {pages}
  {4011} (\bibinfo {year} {1995})}\BibitemShut {NoStop}%
\bibitem [{\citenamefont {Hite}()}]{HiteChat}%
  \BibitemOpen
  \bibfield  {author} {\bibinfo {author} {\bibfnamefont {D.~A.}\ \bibnamefont
  {Hite}},\ }\href@noop {} {}\bibinfo {howpublished} {private
  communication}\BibitemShut {NoStop}%
\end{thebibliography}
\end{document}